\documentclass[aps,prl,twocolumn,showpacs,groupedaddress,floatfix]{revtex4}
\def\FigFactor{0.5}

\usepackage{graphics}

\usepackage{times}

\def\pT{\mbox{$p_T$}}

\def\sqrtsNN{\mbox{$\sqrt{s_\mathrm{NN}}$}}
\def\sqrts{\mbox{$\sqrt{s}$}}
\def\Npart{\mbox{$\mathrm{N}_\mathrm{part}$}}
\def\NpartMean{\mbox{$\langle\Npart\rangle$}}
\def\Nbinary{\mbox{$\mathrm{N}_\mathrm{bin}$}}
\def\NbinaryMean{\mbox{$\langle\Nbinary\rangle$}}

\def\pizero{\mbox{$\pi^0$}}
\def\kzeros{\mbox{$K^0_s$}}
\def\pbar{\mbox{$\bar\mathrm{p}$}}
\def\lam{\mbox{$\Lambda$}}
\def\lambar{\mbox{$\bar\Lambda$}}

\def\pbarp{\mbox{$\pbar+\mathrm{p}$}}

\def\Nch{\mbox{$N_{ch}$}}
\def\dNchdeta{\mbox{$\frac{dN_{ch}}{d\eta}$}}
\def\dNchdetaPerNpart{\mbox{$\frac{2}{\langle\mathrm{N}_\mathrm{part}\rangle}\dNchdeta$}}

\def\GeVc{\mbox{$\mathrm{GeV}/c$}}

\def\hplus{\mbox{$\mathrm{h}^+$}}
\def\hminus{\mbox{$\mathrm{h}^-$}}
\def\hphm{\mbox{$(\hplus+\hminus)/2$}}

\def\meanpT{\mbox{$\lt\pT\gt$}}

\def\RAA{\mbox{$R_{AA}(\pT)$}}
\def\TAA{\mbox{$T_{AA}$}}

\def\sigmaNNinel{\mbox{$\sigma^{NN}_{inel}$}}
\def\sigmaAAgeom{\mbox{$\sigma^{AuAu}_{geom}$}}

\def\lt{\mbox{$<$}}
\def\gt{\mbox{$>$}}

\begin{document}

\title{Centrality dependence of high \pT\ hadron suppression 
in Au+Au collisions at \sqrtsNN=130 GeV 
}


\author{
C.~Adler$^{11}$, Z.~Ahammed$^{23}$, C.~Allgower$^{12}$, J.~Amonett$^{14}$,
B.D.~Anderson$^{14}$, M.~Anderson$^5$, G.S.~Averichev$^{9}$, 
J.~Balewski$^{12}$, O.~Barannikova$^{9,23}$, L.S.~Barnby$^{14}$
% ...
\\
(STAR Collaboration)
}
% ...
\affiliation{$^5$University of California, Davis, California 95616}
% ...
\affiliation{$^{9}$Laboratory for High Energy (JINR), Dubna, Russia}
% ...
\affiliation{$^{11}$University of Frankfurt, Frankfurt, Germany}
\affiliation{$^{12}$Indiana University, Bloomington, Indiana 47408}
% ...
\affiliation{$^{14}$Kent State University, Kent, Ohio 44242}
% ...
\affiliation{$^{23}$Purdue University, West Lafayette, Indiana 47907}
% ...


\date{\today}

\begin{abstract}
Inclusive transverse momentum distributions of charged hadrons within
$0.2\lt\pT\lt6.0$ \GeVc\ have been measured over a
broad range of centrality for Au+Au collisions at
\sqrtsNN=130 GeV. Hadron yields are suppressed at high \pT\ in 
central collisions relative to peripheral collisions and to a
nucleon-nucleon reference scaled for collision geometry. Peripheral
collisions are not suppressed relative to the nucleon-nucleon
reference. The suppression varies continuously at intermediate
centralities. The results indicate significant nuclear medium effects
on high \pT\ hadron production in heavy ion collisions at high
energy.
\end{abstract}


\maketitle


QCD predicts a phase transition at high energy density from hadronic
matter to a deconfined Quark-Gluon Plasma (QGP) \cite{QGP}. This
transition may be studied in the laboratory through the collision of
heavy ions at ultrarelativistic energies. Partons propagating in a
medium lose energy through gluon bremsstrahlung
\cite{GyulassyPlumer,WangGyulassy,BaierA}, with the magnitude of the
energy loss predicted to depend strongly on the gluon density of the
medium. Measurement of partonic energy loss therefore provides a
unique probe of the density of the medium. 

Analysis of deep inelastic scattering and Drell-Yan pair production
using nuclear targets indicates that the energy loss in cold nuclear
matter is $0.2-0.5$ GeV/fm for quarks with energy greater than 10 GeV
\cite{WangAndWang,Arleo}. Hard scattering of partons in nuclear collisions occurs early in the
evolution of the extended system, thereby probing the phase of highest
density. Energy loss softens the fragmentation of jets, leading to the
suppression of high transverse momentum (high \pT) hadrons in the
final state \cite{XNLeadinghadron}. The PHENIX collaboration has
reported the suppression of charged hadron and \pizero\ production at
high \pT\ in central Au+Au collisions at center-of-mass energy per
nucleon pair
\sqrtsNN=130 GeV, relative both to reference data from nucleon-nucleon
(NN) collisions and to peripheral Au+Au collisions
\cite{PhenixHighpT}. The suppression is in qualitative agreement with predictions of
partonic energy loss in dense matter, though quantitative conclusions
require the understanding of other nuclear effects
\cite{PhenixHighpT}.

This Letter presents a measurement of the inclusive
charged hadron yield \hphm\ within $0.2\lt\pT\lt6.0$ \GeVc,
measured for a broad range of centrality in Au+Au collisions at
\sqrtsNN=130 GeV by the STAR collaboration at RHIC. Suppression of
charged hadron production at high \pT\ in central collisions is
observed, in qualitative agreement with the PHENIX
measurement \cite{PhenixHighpT}. The high precision and wide kinematic
and centrality coverage of the data presented here permit a detailed study of
nuclear medium effects on hadron production from the soft to the hard
scattering regime.

For comparison of spectra from nuclear collisions to an NN
reference, the nuclear modification factor is defined as
\begin{equation}
\label{RAA}
\RAA=\frac{d^2N^{AA}/d{\pT}d\eta}
{\TAA\cdot{d}^2\sigma^{NN}/d{\pT}d{\eta}},
\end{equation}

\noindent
where \TAA=\NbinaryMean/\sigmaNNinel\ accounts for the collision
geometry, averaged over the event centrality class. \NbinaryMean, the
equivalent number of binary NN collisions, is calculated using a
Glauber model. $\RAA$ is less than unity at low \pT\
\cite{STARhminus}. In contrast, the yield for hard processes scales as
\NbinaryMean\ in the absence of nuclear medium effects (\RAA=1), and
effects of the medium may be measured at high \pT\ by the deviation of
\RAA\ from unity. In addition to final state energy loss, \RAA\ may be affected by
initial state multiple scattering \cite{XNLeadinghadron,Papp} and
transverse flow, both of which will enhance hadron production at high
\pT, and by shadowing of nuclear parton distributions. At
significantly lower \sqrts\ than the present study, enhancement of
hadron production at high \pT\ has been observed in p-nucleus
\cite{CroninEffectData} and $\alpha-\alpha$ \cite{ISRCronin}
collisions, as well as central collisions of heavy nuclei
\cite{WA98}. Current estimates of shadowing effects at RHIC energies contain large
uncertainties \cite{XNLeadinghadron,XNShadowing}.


\begin{table}
\caption{
Typical multiplicative correction factors and systematic uncertainties, 
applied to the yields for peripheral and central collisions.
\label{TableCorr}}
\begin{ruledtabular}
\begin{tabular}{|c||c|c||c|c|}
		& \multicolumn{2}{c|}{\pT=2 \GeVc} & \multicolumn{2}{c|}{\pT=5.5 \GeVc} \\ 
									\cline{2-5}
centrality		& 60-80\%	 & 0-5\% & 60-80\% & 0-5\% \\ \hline
Tracking & 1.16$\pm$0.08 & 1.59$\pm$0.16 & 1.14$\pm$0.23 & 1.49$\pm$0.30 \\
Background 	& 0.95$\pm$0.03& 0.90$\pm$0.05 & 0.95$\pm$0.05 & 0.88$\pm$0.12 \\
\pT\ resolution	& 1.00$^{+0}_{-0.01}$ & 0.99 $\pm$0.01 & 0.92$\pm$0.05 & 0.76$\pm$0.10 \\ 
\end{tabular}
\end{ruledtabular}
\end{table} 

The STAR detector is described in \cite{STARNIM}. Data collection
utilized both a minimum bias trigger and a trigger selecting the 10\%
most central events. Charged particle tracks were detected in the Time
Projection Chamber (TPC), with momentum determined from their
curvature in a 0.25 T magnetic field. After event selection cuts, the
central data set contained 320,000 events, while the minimum bias data
set contained 240,000 events. The measured minimum bias distribution,
corrected for vertex finding efficiency at low multiplicity,
corresponds to $94{\pm}2$\% of the Au+Au geometric cross section
\sigmaAAgeom, assumed to be 7.2 barns \cite{Baltz}.

Centrality selection is based on the primary charged particle
multiplicity \Nch\ within the pseudorapidity range $|\eta|\lt0.5$. The
most central bin is 0-5\% of \sigmaAAgeom, while the most peripheral
bin is 60-80\%. Alternative centrality measures incorporate forward
neutral energy \cite{STARhminus,RHICMult} and its correlation with
multiplicity at midrapidity. The maximum variation of the \pT\
spectrum for different centrality measures is 4\% for central events
and less than 4\% for more peripheral events. This variation is
included in the systematic uncertainties of the reported spectra.

Analysis of inclusive charged particle spectra for $\pT\lt2$ \GeVc\
has been described previously \cite{STARhminus}. Accepted tracks for
$\pT\gt2$ \GeVc\ have $|\eta|\lt0.5$ and a distance of closest
approach to the primary vertex less than 1 cm to reject
background. Acceptance, efficiency, and momentum resolution were
determined by embedding simulated tracks into real raw data
events. For \pT\gt1.5 \GeVc, the Gaussian distribution of track
curvature $k\propto{1}/\pT$ has relative width
$\delta{k}/k=0.016\cdot(\pT/(\GeVc))+0.012$ for central events and
$\delta{k}/k=0.011\cdot(\pT/(\GeVc))+0.013$ for peripheral
events. Correction was made for distortion due to finite momentum
resolution. 

The hadron yield decreases rapidly with increasing \pT\ and its
measurement is sensitive to small spatial distortions, which generate
charge sign-dependent systematic errors in the measured track
curvature. Measurement of the summed hadron yield,
\hphm, is markedly less sensitive to such distortions than the yield
of one charge sign alone. Each half of the cylindrical TPC is divided
azimuthally into 12 sectors. High \pT\ tracks have small sagitta
($s\sim0.8$ cm at \pT=5 \GeVc) and are confined to a single
sector. The sector-wise distribution of \hphm\ has a
\pT-dependent rms variation of less than 5\%, though with 
correlated variations for adjacent sectors. These effects contribute to
the systematic uncertainty.

The most significant background corrections are due to weak particle decays
and anti-nucleon annihilation in detector material. The former are
estimated based on \lam, \lambar\ and \kzeros\ yields measured for
$\pT<2.5$ \GeVc\
\cite{STARLambda,STARkzero}, with extrapolation to higher \pT\ 
using an exponential fit \cite{STARLambda}.  The latter are based on
measured anti-proton yields
\cite{STARPHENIXpbar}.

The major correction factors and their uncertainties are given in Table
\ref{TableCorr}. ``Tracking'' incorporates efficiency, acceptance, and 
the effects of the spatial non-uniformity of the TPC, with the latter
dominating its systematic uncertainty. The total systematic
uncertainty of the spectra is the quadrature sum of the uncertainties
in Table \ref{TableCorr}. For the highest
\pT\ bins it is $\approx 27\%$ for central events and $\approx 21\%$
for peripheral events.


\def\TableTwoCaption{Geometric quantities, charged particle density per participant pair, 
and fit parameters to Eq. (\ref{PowerLaw}), for various centrality bins and
for \pbarp\ at 200 GeV \cite{UA1}, assuming \sigmaNNinel=41 mb. The
fits to the Au+Au data use uncorrelated measurement errors, which
are largely systematic and non-Gaussian. Parameter errors shown also
include correlated systematic uncertainties, which are added to
parameter errors resulting from the fit.
}

\begin{table*}
\caption{\TableTwoCaption
\label{TableTwo}}
\begin{ruledtabular}
\begin{tabular}{|c||c|c||c|c||c|c|c||}
 & & & & & \multicolumn{3}{c||}{Power Law Fit ($0.2<\pT<6.0$ \GeVc)} \\ \cline{6-8}
centrality  & \NbinaryMean & \NpartMean & \dNchdeta & \dNchdetaPerNpart
          & $C ((\GeVc)^{-2})$ & $<\!\pT\!> (\GeVc)$ & $n$  \\ \hline
0-5\%  & 965$^{+67}_{-67}$ & 350$^{+4}_{-4}$ & 563$\pm$39 & 3.22$\pm$0.23
             & 797$\pm$60 & 0.520$\pm$0.010 & 21.9$\pm$0.5 \\
5-10\%  & 764$^{+59}_{-63}$ & 296$^{+7}_{-7}$ & 452$\pm$32 & 3.05$\pm$0.23
             & 654$\pm$50 & 0.517$\pm$0.010 & 20.7$\pm$0.4 \\
10-20\%  & 551$^{+48}_{-56}$ & 232$^{+9}_{-9}$ & 344$\pm$24 & 2.96$\pm$0.24
             & 520$\pm$40 & 0.511$\pm$0.010 & 18.9$\pm$0.4 \\
20-30\%  & 348$^{+44}_{-45}$ & 165$^{+10}_{-10}$ & 234$\pm$16 & 2.83$\pm$0.26
             & 372$\pm$28 & 0.504$\pm$0.011 & 17.3$\pm$0.3 \\
30-40\%  & 210$^{+36}_{-36}$ & 115$^{+10}_{-12}$ & 144$\pm$10 & 2.51$\pm$0.30
             & 252$\pm$19 & 0.497$\pm$0.011 & 17.2$\pm$0.3 \\
40-60\%  & 90$^{+22}_{-22}$ & 62$^{+9}_{-11}$ & 72.2$\pm$5.1 & 2.35$\pm$0.42
             & 141$\pm$11 & 0.480$\pm$0.011 & 14.8$\pm$0.2 \\
60-80\%  & 20$^{+7}_{-9}$ & 20$^{+5}_{-6}$ & 21.0$\pm$1.5 & 2.13$\pm$0.61
             & 50$\pm$4 & 0.446$\pm$0.011 & 13.0$\pm$0.2 \\
\hline \pbarp(200) & 1 & 2 & $2.65\pm0.08$ & $2.65\pm0.08$ & 7.7$\pm$0.4 & 0.392$\pm$0.003 & 11.8$\pm$0.4 \\
\end{tabular}
\end{ruledtabular}
\end{table*}


\begin{figure}
\resizebox{\FigFactor\textwidth}{!}{\includegraphics{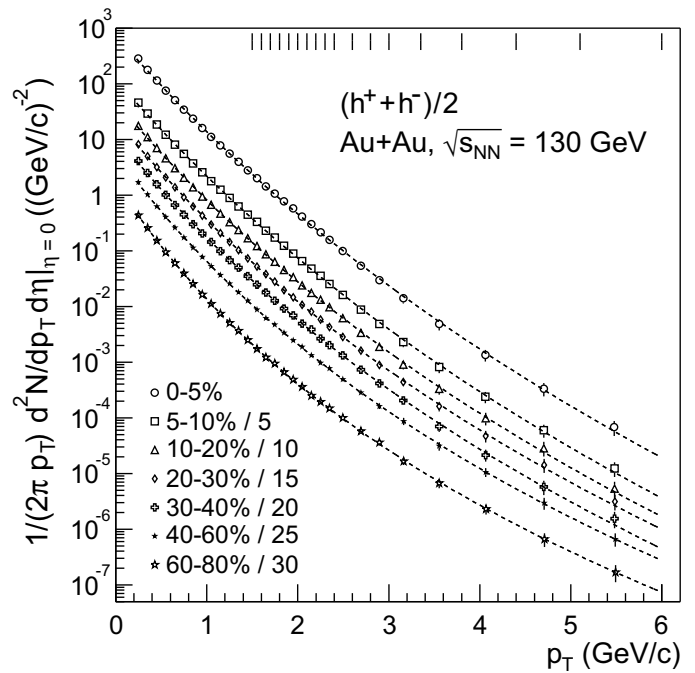}}
\caption{Inclusive \pT\ distributions of \hphm. Non-central bins are scaled down 
by the indicated factors. The combined statistical and systematic
errors are shown. Curves are fits to Eq. (\ref{PowerLaw}). Hash
marks at the top indicate bin boundaries for \pT\gt1.5
\GeVc.\label{FigOne}}
\end{figure}

Figure \ref{FigOne} shows the inclusive \pT\ distributions of
\hphm\ within $|\eta|\lt0.5$ for various centrality bins, for Au+Au
collisions at \sqrtsNN=130 GeV. Error bars, which are dominated by
systematic uncertainties at all \pT, are generally smaller than the
symbols. The spectra were fit by the pQCD-inspired power law function
\cite{UA1}

\begin{equation}
\label{PowerLaw}
\frac{1}{2\pi\pT}\frac{dN}{d\pT}=C\cdot(1+\pT/p_0)^{-n},
\end{equation}

\noindent
which describes \pT\ spectra of charged hadrons for NN collisions over
a wide range of \sqrts\ \cite{ISR,UA1,CDF}. Systematic changes in
shape of the spectra with centrality are revealed by the fit
parameters $C$, $n$, and mean transverse momentum
\meanpT=$2p_0/(n-3)$ in Table
\ref{TableTwo}. Also shown in the Table are fit parameters for
\pbarp\ collisions at \sqrts=200 GeV \cite{UA1}.
Eq.~(\ref{PowerLaw}) yields a poor fit relative to the uncertainties of the data for the
most central bin, with systematic deviations from the fit function of
10-20\%, whereas for more peripheral collisions it fits well, with
parameters tending smoothly to those for
\pbarp\ collisions.

A more direct comparison of yields for different centralities relies
on estimating \NbinaryMean\ and the mean number of participants
\NpartMean\ for each centrality bin. For this purpose, the
distribution $d\sigma/d\Nbinary$ (and equivalently, $d\sigma/d\Npart$)
was calculated using a Monte Carlo Glauber model \cite{RHICMult} with
\sigmaNNinel=$41\pm1$ mb and Woods-Saxon nuclear matter density, using
radius $r=6.5\pm0.1$ fm and surface diffuseness $a=0.535\pm0.027$
fm. \cite{Baltz}. Percentile intervals of $d\sigma/d\Nbinary$
(calculated) and $d\sigma/d\Nch$ (measured) were equated to extract
\NbinaryMean\ for each centrality bin. The Glauber model parameters and 
geometric cross section were varied to estimate the systematic
uncertainties. Results are given in Table
\ref{TableTwo}.

Because charged particle distributions at midrapidity are used for
centrality selection, biases in the relation between
$d\sigma/d\Nbinary$ and $d\sigma/d\Nch$ due to fluctuations and
autocorrelations were assessed. Variation of parameters in a Monte
Carlo calculation of multiplicity fluctuations for fixed collision
geometry, and comparison of the measured multiplicity distribution in
an azimuthal quadrant centered on a high
\pT\ particle against those in the other quadrants, both yielded 
negligible uncertainties in \NbinaryMean\ and \NpartMean.

Table \ref{TableTwo} shows the charged particle yield per participant
pair, \dNchdetaPerNpart, obtained by integrating the \pT\ spectra. The
extrapolated yield in $\pT\lt0.2$ \GeVc\ is $\sim20$\% of the
total for all centralities. The dependence of
\dNchdetaPerNpart\ on
\NpartMean\ is consistent with observations
in \cite{RHICMult}.

\begin{figure}
\resizebox{\FigFactor\textwidth}{!}{\includegraphics{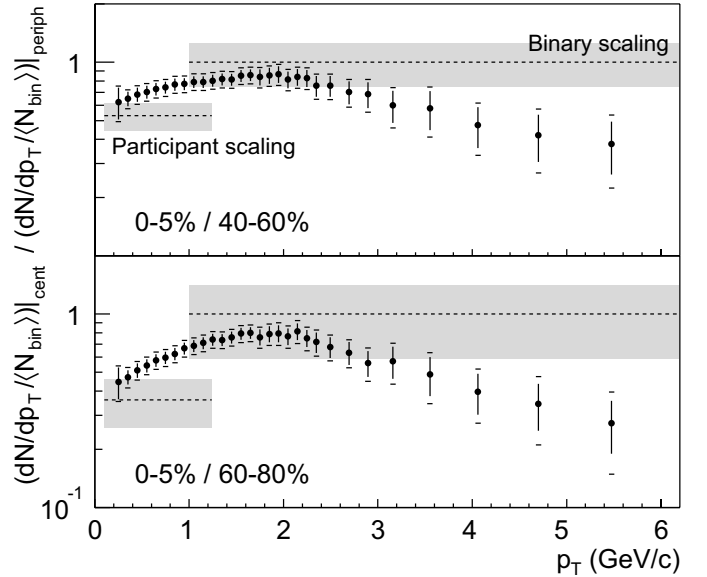}}
\caption{Ratio of charged hadron yields within $|\eta|<0.5$ for central 
over peripheral collisions, normalized to \NbinaryMean.
\label{FigTwo}}
\end{figure}

Figure \ref{FigTwo} shows the ratio of the central (0-5\%) spectrum to
that of the two peripheral bins (40-60\%, 60-80\%), normalized by
\NbinaryMean. The dashed lines at unity and below show scaling with
\NbinaryMean\ and \NpartMean\ respectively, and the shaded regions show the systematic uncertainties
from Table \ref{TableTwo}.  The vertical error bars on the data points
are the uncertainties of the central data, while the horizontal caps are the
quadrature sum of the uncertainties of both data sets. Approximate
participant scaling at low \pT\ is seen. The ratio rises monotonically
below $\pT\sim2$ \GeVc\ and decreases at high \pT. The ratio of central over
most peripheral achieves a value at
\pT=5.5 \GeVc\ of $0.27{\pm}0.12$ with additional uncertainty $\pm0.12$ due to \NbinaryMean, 
establishing significant suppression of charged hadron production at
high \pT\ in central collisions.

\begin{figure}
\resizebox{\FigFactor\textwidth}{!}{\includegraphics{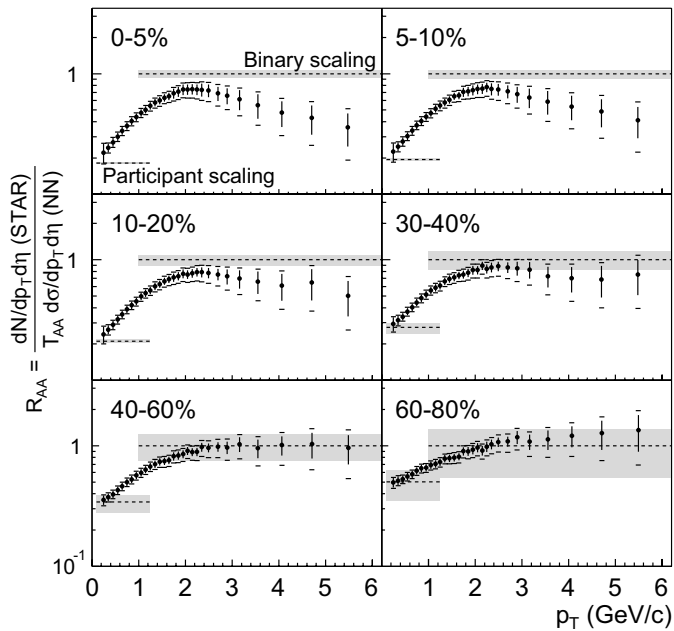}}
\caption{\RAA\ for various centrality bins, for Au+Au relative to an NN reference spectrum. 
Error bars are described in the text. Errors between different \pT\ and
centrality bins are highly correlated.
\label{FigThree}}
\end{figure}

Figure \ref{FigThree} shows \RAA\ for various centrality bins relative
to an NN reference spectrum, parameterized by Eq.~(\ref{PowerLaw}) with
$C\sigmaNNinel=267^{+4}_{-6}$ mb/(\GeVc)$^2$, $p_0=1.90^{+0.17}_{-0.09}$
\GeVc, and $n=12.98^{+0.92}_{-0.47}$ (the superscripts and subscripts
are curves that bound the systematic uncertainty).  The
reference was determined by fitting Eq. (\ref{PowerLaw}) to UA1
\pbarp\ data at \sqrts=200-900 GeV \cite{UA1} and extrapolating
to \sqrts=130 GeV. Extrapolation of the UA1 200 GeV spectrum to 130 GeV
using pQCD calculations agrees to within 5\% with the reference at
\pT=6 \GeVc\ \cite{Pythia,IvanInterp}. Correction to the NN reference
for the UA1 acceptance ($|\eta|\lt2.5$), which differs from this analysis
($|\eta|\lt0.5$), was based on two independent pQCD calculations
\cite{Pythia,Yuan}, giving a multiplicative correction of
$1.17\pm0.06$ at \pT=2.0 \GeVc\ and $1.35\pm0.10$ at
\pT=5.5 \GeVc. The \pT-dependent systematic uncertainty of the NN reference is the quadrature
sum of the power law parameter and the acceptance correction
uncertainties. Isospin effects are negligible for \pT$\lt$6
\GeVc\ \cite{IvanInterp}. The error bars are the systematic
uncertainties of the measured spectra, while the caps show their
quadrature sum with the systematic uncertainty of the NN reference.

\RAA\ increases monotonically 
for $\pT\lt2$ \GeVc\ at all centralities, and saturates near unity for
$\pT\gt2$ \GeVc\ in the most peripheral bins. In contrast, \RAA\ for the central
bins reaches a maximum and then decreases strongly above \pT=2 \GeVc, showing
suppression of the charged hadron yield relative to the NN reference of
$0.36\pm0.16$ (sys) at \pT=5.5 \GeVc\ for the 0-5\% bin, with
additional uncertainty $\pm0.03$ due to
\NbinaryMean. \RAA\ varies continuously as a function of
centrality, and no centrality threshold for the onset of suppression is
observed \cite{XNWThresholdSuppression}.


In summary, charged hadron production in high energy collisions of
heavy nuclei has been studied over a wide range of \pT\ and event
centrality. At high \pT, hadron yields scale with the number of binary
collisions for peripheral collisions, while significant suppression of
hadron production is seen for central collisions. This phenomenon
indicates substantial energy loss of the final state partons or their
hadronic fragments in the medium generated by high energy nuclear
collisions, though quantitative measurement of this effect requires
additional reference data.


\begin{acknowledgments}
We thank the RHIC Operations Group and the RHIC Computing
Facility at Brookhaven National Laboratory, and the National Energy
Research Scientific Computing Center at Lawrence Berkeley National
Laboratory for their support. This work was supported by the Division
of Nuclear Physics and the Division of High Energy Physics of the
Office of Science of the U.S. Department of Energy, the United States
National Science Foundation, the Bundesministerium f\"ur Bildung und
Forschung of Germany, the Institut National de la Physique Nucl\'eaire
et de la Physique des Particules of France, the United Kingdom
Engineering and Physical Sciences Research Council, Funda\c c\~ao de
Amparo \`a Pesquisa do Estado de S\~ao Paulo, Brazil, the Russian
Ministry of Science and Technology and the Ministry of Education of
China and the National Science Foundation of China.
\end{acknowledgments}


\def\etal{\mbox{$\mathrm{\it et\ al.}$}} 
\def\QM{Proceedings of Quark Matter 2001, Stony Brook, N.Y., 
Jan. 15-21, 2001. Nucl.\ Phys.\ A, in print}


\end{document}